\documentclass[aps,prb,twocolumn,showpacs,superscriptaddress]{revtex4}
\usepackage{epsfig}
\usepackage{amsmath}
\usepackage[dvipsnames,usenames]{color}
\emergencystretch=\maxdimen
\begin{document}

\title{Superconducting Transitions in Flat Band Systems}

\author{V.I. Iglovikov}
\affiliation{Physics Department, University of California, Davis,
California 95616, USA}

\author{F. H\'ebert}
\affiliation{INLN, Universit\'e de Nice--Sophia Antipolis, CNRS;
1361 route des Lucioles, 06560 Valbonne, France}

\author{B. Gr\'emaud}
\affiliation{Laboratoire Kastler Brossel, Ecole Normale Sup\'{e}rieure
  CNRS, UPMC; 4 Place Jussieu, 75005 Paris, France} 
\affiliation{Centre for Quantum Technologies, National University of
  Singapore, 3 Science Drive 2, Singapore 117543, Singapore} 
\affiliation{Department of Physics, National University of Singapore,
  2 Science Drive 3, Singapore 117542, Singapore} 
 
\author{G.G. Batrouni}
\affiliation{INLN, Universit\'e de Nice--Sophia Antipolis, CNRS;
1361 route des Lucioles, 06560 Valbonne, France}
\affiliation{Centre for Quantum Technologies, National University of
  Singapore, 3 Science Drive 2, Singapore 117543, Singapore} 
\affiliation{Institut Universitaire de France, 103, Boulevard
  Saint-Michel, 75005 Paris, France}

\author{R.T.  Scalettar}
\affiliation{Physics Department, University of California, Davis,
California 95616, USA}

\begin{abstract}
The physics of strongly correlated quantum particles within a flat
band was originally explored as a route to itinerant ferromagnetism
and, indeed, a celebrated theorem by Lieb rigorously establishes
that the ground state of the repulsive Hubbard model on a bipartite
lattice with unequal number of sites in each sublattice must have
nonzero spin $S$ at half-filling.  
Recently, there has been interest in Lieb geometries due to the possibility
of novel topological insulator, nematic, and Bose-Einstein condensed
(BEC) phases.  In this paper, we extend the understanding of the
attractive Hubbard model on the Lieb lattice by using Determinant
Quantum Monte Carlo to study real space charge and pair correlation
functions not addressed by the Lieb theorems.  
Specifically, our results show unusual charge 
and charge transfer signatures
within the flat band, and a reduction in pairing order at
$\rho=2/3$ and $\rho=4/3$, the points at which
the flat band is first occupied and then completely filled.
We compare our results to the case of flat bands in
the Kagome lattice and
demonstrate that the behavior observed in the two cases is rather 
different.
\end{abstract}


\maketitle

\section{Introduction}

The form of the electronic dispersion
relation $\epsilon(k)$ in the absence of interactions plays a
fundamental role in how correlations drive the formation of ordered phases.  
Qualitative pictures like the Stoner criterion for
ferromagnetism simplify the input from $\epsilon(k)$ and
focus on the density of states at the Fermi
level $N(E_F) = \sum_k \delta(\, E_F - \epsilon(k) \,)$.  
More refined treatments like the Random Phase Approximation (RPA)
capture phenomena such as the degree of Fermi
surface nesting and 
provide insight into how the
non-interacting susceptibility determines the renormalized response of
the system.  Both density of states and nesting issues 
come into play in 
cuprate superconductity: Near-neighbor hopping on a two dimensional
square lattice such as that occupied by the copper atoms of the
CuO$_2$ sheets has a van Hove singularity in the density of states at
half-filling which was suggested to lead to an enhanced
superconducting critical temperature\cite{hirsch86}.  Likewise nesting
of the Fermi-surface with wave vector ${\bf q} = (\pi,\pi)$ provides a
natural weak-coupling explanation for the antiferromagnetic phase of
 the undoped parent compounds, complementing the strong-coupling
 Heisenberg picture.  Nesting can also further increase the pairing
 transition temperature\cite{scalapino87}.
%% v20 the undoped parent compounds and can increase $T_c$\cite{scalapino87}.

\begin{figure}[t] 
\epsfig{figure=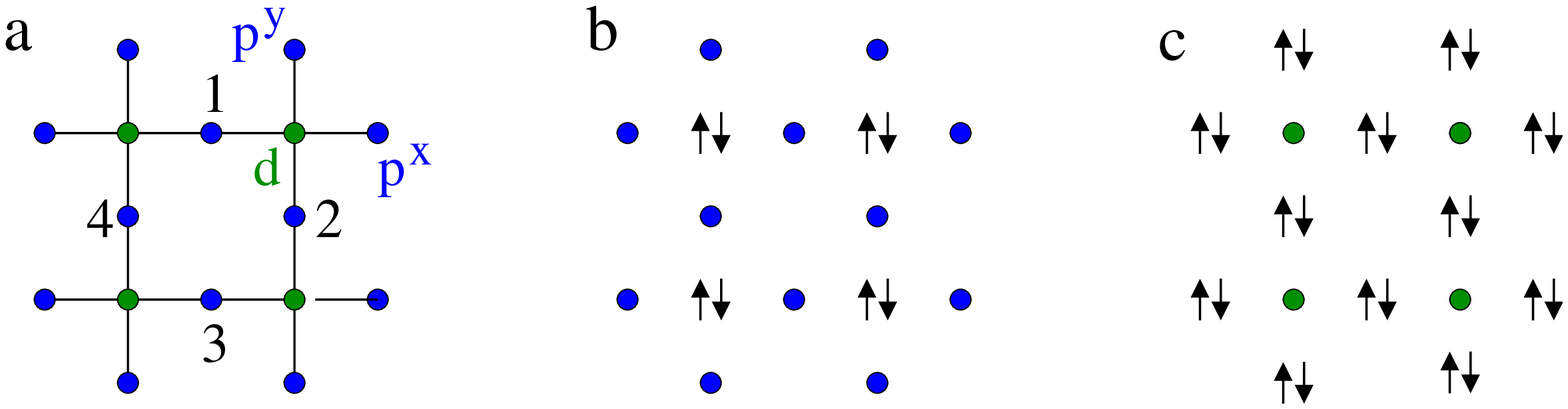,width=8.5cm,angle=0,clip}
\caption{(Color online) 
An example of a Lieb lattice, a bipartite
  geometry with unequal numbers of sites in the two sublattices.  (a)
The three bands model of the CuO$_2$ planes of the cuprate.
In the case of attractive interactions CDW patterns
  emerge at two-thirds (b) and four-thirds 
  (c) fillings
 by doubly occupying the copper or oxygen sublattices respectively.  
Localized states form from linear
    combinations of creation operators on sites 1,2,3,4 (panel a) with
    alternating phases.  
\label{fig:cdwpatterns2}
}
\end{figure}

While the single band Hubbard model on a square lattice has received
the most attention in modeling the cuprates, considerable interest has
also focussed on the more accurate three band picture which includes
not only the square lattice of copper $d$ orbitals but also the
intervening oxygen $p$ orbitals
\cite{emery87,varma87,zaanen85,oles89,scalettar91,fischer11}.  If
hopping is restricted to near-neighbors,
this arrangement of sites is bipartite, with, however, unequal numbers
$N_p = 2 N_d$.  In such situations, Lieb showed\cite{lieb89} that, at
half-filling and with repulsive interaction, the hopping Hamiltonian
$\hat T$ has a ground state with non-zero spin, $S=(N_p - N_d)/2$.  
The key element of the physics of such `Lieb lattices' is that the
spectrum of 
$\hat T$ consists of $2N_d$
eigenvalues in +/- pairs, separated by a flat electronic band
$\epsilon(k)=0$ with $N_p-N_d$ levels.  Fig.~\ref{fig:cdwpatterns2}a
shows an example of a Lieb geometry ($N_p=2N_d$). 
While this structure is similar
to the CuO$_2$ planes of the high temperature superconductors,
in a realistic cuprate model there is an
energy difference between the copper $d$ and oxygen $p$ orbitals.  The zero
energy modes of $\hat T$ can be easily understood: A one particle
state formed by creating fermions on the four oxygen sites surrounding
the center of any copper plaquette, $|\psi\rangle = (c_1^\dagger
-c_2^\dagger +c_3^\dagger -c_4^\dagger ) | {\rm vac} \rangle$,
satisfies $\hat T | \psi \rangle = 0$ because hops from the oxygens
onto the coppers cancel.
This `topological' localization was emphasized earlier by
Sutherland \cite{sutherland86}.

The Lieb lattice, as realized in CuO$_2$ planes of the cuprate
superconductors, was investigated by Varma for possible staggered
current phases which might explain pseudogap behavior \cite{varma06}.
%v20 The Lieb lattice realized in CuO$_2$ planes was investigated for staggered
%v20 current phases and pseudogap physics \cite{varma06}.
Novel behavior tied to the presence of a flat band includes
a quantum spin Hall effect driven by spin-orbit coupling and
topological phase transitions, e.g.~caused by next-near-neighbor
hopping \cite{everts12}.  Other `decorated' geometries exhibit flat
bands, e.g.~the Kagome lattice, and are similarly under investigation
\cite{kagome1,kagome2}, a key difference being the frustrated nature of the Kagome lattice. 
We compare results obtained for the Lieb lattice to
the Kagome lattice and show the results we find are not generic to
arbitrary flat band systems.

In this paper we address two important questions left open by Lieb's
theorems: [1] What are the natures of the charge and superconducting
response functions?  [2] What are the implications of the absence of a
minimum in $\epsilon(k)$ for superconductivity (SC) or BEC in a flat
band?  It has been suggested that the presence of interactions
renormalizes the flat band and induces an effective minimum so that
BEC can still occur\cite{huber10} or that the infinite density of
states in the flat band favours the emergence of SC or other kinds of
order\cite{kopnin11,kopnin13,szhang2010}.  However no exact numerical
work has addressed these issues.  Next-generation Optical Lattice
Emulation (OLE) experiments have generated Lieb lattice
geometries\cite{apaja10,shen10} and might be able to study this
question for bosonic atoms \cite{stamperkurn13}.  
The attractive fermion
Hubbard model (AHM) considered here 
develops superconducting phases at low temperatures- a 
 BCS phase at weak coupling and BEC pairing at strong
  coupling\cite{randeria13}, with a crossover between these two
  extreme cases.  On lattices which do not have
special features in their densities of states\cite{kopnin11,kopnin13},
and in sufficiently high dimension, the BCS limit is characterized by
pairs with large sizes $\xi$ and a transition temperature $T_c \sim t
\, {\rm exp}[\,-a t/|U|] \sim t \, {\rm exp} [\,-b / U N(E_{\rm
  fermi}) \,]$.  In the BEC limit, $\xi$ is of the order of a few
lattice spacings and $T_c \sim t^2/|U|$, the effective hopping of the
tightly bound pairs. We will concentrate on intermediate and large
coupling cases ($|U| \ge 4$).  This is
closer to the BEC limit, and hence to possible
experiments on bosonic atoms.  Furthermore, it is
easier to reach the condensation temperature in this case.

%% \begin{figure}[h!] 

\section{Calculational Approach} 
We consider the AHM on a CuO$_2$
geometry (Fig.~\ref{fig:cdwpatterns2}a).
\begin{eqnarray}
\label{eq:ham}
H &=& -t \sum_{i \alpha\sigma}
\big( \,
d_{i \sigma}^{\dagger}   \, p_{i \sigma}^{\, \alpha} 
+ d_{i+ \alpha \, \sigma}^{\dagger} \, 
p_{i \sigma}^{\, \alpha} 
+{\rm h.c.} \, \big)
 \\
 &&- |U| \sum_{i\alpha} 
 \biggl[(\, n_{i\uparrow}^{d} -1/2 \, )
 (\, n_{i \downarrow }^{d} -1/2 \, ) \\
 &&\hspace{1.7cm} +
 (\, n_{i \uparrow}^{p \alpha} -1/2 \, )
 (\, n_{i  \downarrow }^{p\alpha} -1/2 \, ) \biggr]
\nonumber
\end{eqnarray}
We have adopted the notation of the three band model of the cuprates
where the operators $d_{i \sigma}^{\dagger} \,(d_{i
  \sigma}^{\phantom{\dagger}})$ create (destroy) fermions on site $i$
of spin $\sigma$ in a square lattice of copper $d$ orbitals and $p_{i
  \sigma}^{ \alpha \dagger} \, (p_{i \sigma}^{ \alpha
  \phantom{\dagger}})$ do the same for oxygen $p$ orbitals on the
intervening links in the $\alpha =\hat x,\hat y$ directions.  Number
operators are denoted by $n_{i\sigma}^{p \alpha}$ and
$n_{i\sigma}^{d}$.  $t$ is the scale of kinetic energy which we set to
unity and $|U|$ is the magnitude of the on-site attraction.
%% \cite{footnote1}.  
$N$ denotes the total number of sites of
the lattice \cite{footnote2}.

\begin{figure}[t] 
\epsfig{figure=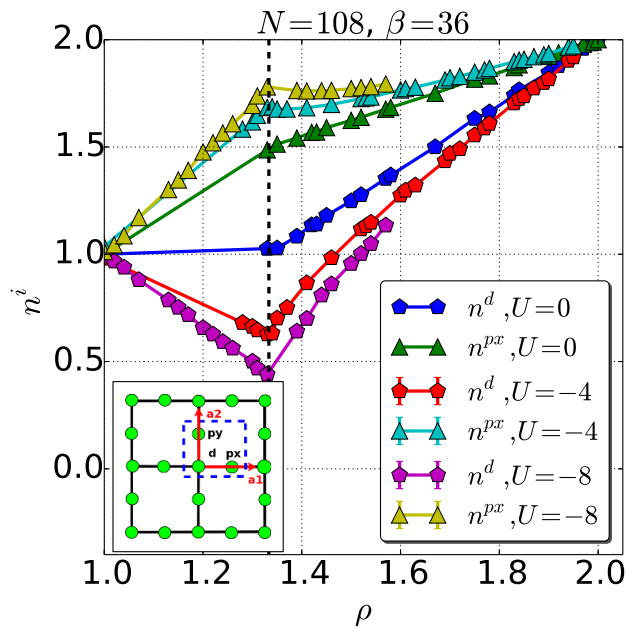,width=8.5cm,angle=0,clip}
\caption{(Color online) Occupations of the copper ($n^d$) and oxygen
  ($n^{p\alpha}$)  orbitals as a function of density $\rho$ for $U=0,-4,-8$.
Non-zero attractive interactions induce an unusual 
charge transfer effect in which the copper occupation decreases even while the
overall lattice density increases, with cusps at the endpoint
of the filling of the flat band, $\rho=4/3$.
Here and in subsequent figures we
show only densities $\rho \geq 1$ since our model Eq.~\ref{eq:ham} is
particle-hole symmetric. 
\label{Fig:n0_n1_vs_rho}
}
\end{figure}

In order to determine the properties of the AHM on a Lieb lattice,
Eq.~\ref{eq:ham}, we use Determinant Quantum Monte Carlo (DQMC)
\cite{blankenbecler81,white89}.  The approach exactly solves the
Hamiltonian on lattices of finite size \cite{footnote3}.  We present
results for up to 6$\times$6 unit cells (108 sites).  The absence of
the sign problem in the attractive case, $U<0$, allows 
simulations over a wide range of fillings.  We focus our attention on
the densities on the $d$ and $p$ orbitals, $n^d$ and $n^{px}=n^{py}$,
and on intra-unit-cell (ie near-neighbor and next-near neighbor)
density-density correlations, $\langle n^d n^{px} \rangle$ and
$\langle n^{px} n^{py} \rangle$.  The total density per site
$\rho=\frac{1}{3}(n^d + n^{px}+n^{py})$.  We also report data for the
local moment $\langle m^2 \rangle = \sum_{i\nu} \langle \, (
\,n_{i\nu\uparrow} - n_{i\nu\downarrow})^2 \,\rangle $ and s-wave pair
structure factor $P_s = 1/N^2 \,\, \sum_{i,j,\mu,\nu} \langle
B^{\phantom{\dagger}}_{i\mu} B^{\dagger}_{j\nu} \rangle$.  Here
$B^{\dagger}_{j\nu}$ creates a pair of up/down spin fermions on site
$j$ and orbital $\nu=d,p^x,p^y$. In the definitions of $\langle m^2
\rangle$ and $P_s$ the sums are over all $N$ lattice sites (i.e.~over
both $d$ and $p$ orbitals).

We complement these DQMC calculation with mean field theory (MFT).  In
this approach, on each site, the attractive interaction is written in
term of the operators
%% \begin{eqnarray}
\begin{align}
  \Delta_x&=\frac{1}{2}(c_{\uparrow}^{\dagger}
  c_{\downarrow}^{\dagger} +c_{\downarrow}^{\phantom{\dagger}}
  c_{\uparrow}^{\phantom{\dagger}}) \hskip0.50in
  \Delta_y=\frac{1}{2i}(c_{\uparrow}^{\dagger}
  c_{\downarrow}^{\dagger}-c_{\downarrow}^{\phantom{\dagger}}
  c_{\uparrow}^{\phantom{\dagger}})
  \nonumber \\
  \Delta_z&=\frac{1}{2}(n_{\uparrow}^{\phantom{\dagger}}
  +n_{\downarrow}^{\phantom{\dagger}}-1),
 \\ 
-|U|&(n_{\uparrow}-1/2)(n_{\downarrow}-1/2) =
-2|U|\vec{\Delta}\cdot\vec{\Delta}/3+|U|/4.  
  \nonumber
\end{align}
%% \end{eqnarray}
The vector $\vec{\Delta}=(\Delta_x,\Delta_y,\Delta_z)$ is obtained
from the usual spin operator by a particle-hole transformation for the
down spin. The mean field decoupling corresponds to approximating, on
each site, the interaction term $\vec{\Delta}\cdot\vec{\Delta}$ by
$2\langle\vec{\Delta}\rangle\cdot\vec{\Delta}$, leading to three mean
field parameters $\langle\vec{\Delta}\rangle$.  For repulsive
interaction, this decoupling is the analog of the usual SU(2)
decoupling in the spin channels~\cite{Minar2013}.  The values
$\langle\vec{\Delta}_i\rangle$ are determined by minimizing the free
energy.

Both techniques work in the grand canonical ensemble, which could cause
a problem when trying to access a given density in the partially filled
flat band. This is not the case as the flat band acquires a width due to
interactions in the exact DQMC treatment and as the SU(2) symmetry
allows us to circumvent this problem in the MF approximation (see
below).

\section{Numerical Results} 
\subsection{Local density response}
We begin by showing the changes in
the occupations $\langle \, n^d \, \rangle$ and $\langle \, n^{p x} \,
\rangle= \langle \, n^{p y} \, \rangle$ on the individual orbitals as
the total density $\rho$ increases (Fig.~\ref{Fig:n0_n1_vs_rho}).  
In the noninteracting limit, $\partial \langle \, n^\alpha \, \rangle /
\partial \rho$ must be nonnegative, and we observe this.  However, for
$U \neq 0$ we see that the $d$ occupation decreases with increasing
$\rho$.  We interpret this in terms of a transition from the
$\rho=2/3$ CDW (Fig.~\ref{fig:cdwpatterns2}b) to the $\rho=4/3$ CDW
(Fig.~\ref{fig:cdwpatterns2}c).  $U$ favors doubly occupied sites, but
to second order in perturbation theory such paired sites are lower in
energy by $2 t^2/|U|$ for each empty adjacent site.  At low densities
(double) occupation of the four-fold coordinated $d$ orbitals is
favored, but as $\rho$ exceeds 2/3 it becomes advantageous to occupy
the more numerous $p$ orbitals and empty the $d$ band.  

We emphasize an important feature of CDW patterns on this lattice:
Because the $d$ and $p$ sites are inequivalent, there is a `trivial'
difference in charge densities which does not reflect any spontaneous
symmetry breaking.  However, CDW order is also present due to
correlations, and the fact that it is energetically favorable, $\delta^2
E \sim t^2/U$, to have doubly occupied and empty sites adjacent.

Additional interest in the orbital occupation evolution
concerns its possible implications
for charge transfer processes in cuprate pairing.  Because they favor
$d$ wave pairing symmetry, it is generally accepted that spin
fluctuations provide the majority of the `pairing glue' in high $T_c$
materials \cite{scalapino94}.  However, arguments have been made
\cite{emery87,varma87,zaanen85,oles89,scalettar91,fischer11} in favor
of a possible role of charge fluctuations between the copper and
oxygen orbitals driven by a repulsive interaction $V_{pd}$.  Such
fluctuations would be reflected in a large response of $n_p - n_d$ to
the orbital energy difference $\epsilon_p - \epsilon_d$.  Figure
\ref{Fig:n0_n1_vs_rho} emphasizes that, even in the absence of
$V_{pd}$, there is nontrivial structure in the orbital occupations.
Thus, much as the large $U=0$ antiferromagnetic susceptibility
highlights spin fluctuations on the
square lattice, the observation of unusual charge transfer in the Lieb
lattice at $V_{pd}=0$ might indicate a role for
charge fluctuations there.

\begin{figure}[t] 
\epsfig{figure=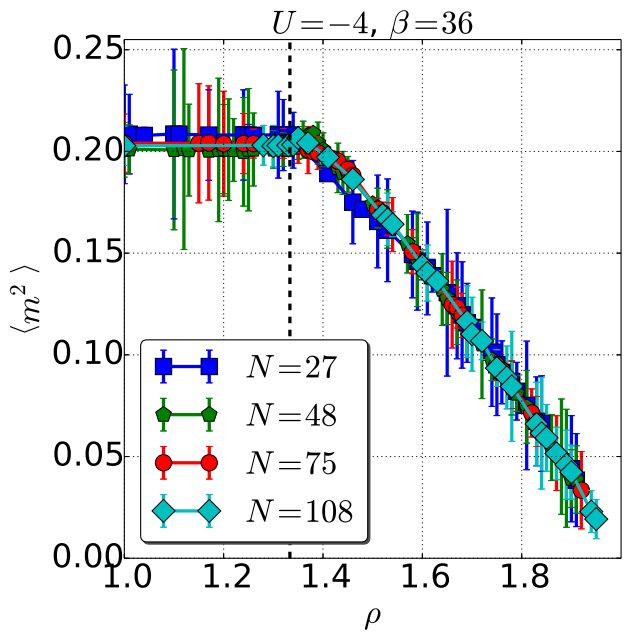,width=8.5cm,angle=0,clip}
\caption{(Color online) The local magnetic moment $\langle m^2
  \rangle$ versus density.  $\langle m^2 \rangle$ is constant within
  the flat band, and then drops for densities $\rho > 4/3$.  Here
  $U=-4t$ and $\beta=36$.  The constant value of $\langle m^2 \rangle$
  can be explained within MFT to be a consequence of the SU(2) invariance at
  $\mu=0$, see the text for more details.  }
\label{Fig:m2_vs_rho_u-4}
\end{figure}

Because it probes the
double occupancy $D$ (local pair formation) the local moment can also
provide interesting insight.
$\langle m^2 \rangle =
\rho - 2 D$ is shown in Fig.~\ref{Fig:m2_vs_rho_u-4}.  It is evident
that $\langle m^2 \rangle$ does not change as the density is increased
in the range $2/3 < \rho < 4/3$, {\it i.e.}~as the flat band is being
filled.  This result can be understood within MFT: for chemical
potential $\mu=0$, the SU(2) invariance implies that the mean field
ground state energy is invariant under a global rotation of the mean
field pseudo-spins $\langle\vec{\Delta}_i\rangle$.  More precisely, in
the ground state, $\langle\vec{\Delta}_i\rangle$ shows a ferromagnetic
order in the $(X,Y)$ plane (pairing order) and an antiferromagnetic
order along the $Z$ axis (CDW order).  
This SU(2) symmetry implies,
\begin{equation}
 \frac{P_s}{a}+\left(\frac{\rho-1}{b}\right)^2=1,
\end{equation}
where $9a=(2\Delta_p+\Delta_d)^2$ and $3b=4\Delta_p-2\Delta_d$, and
$\Delta_i = |\langle \vec{\Delta}_i\rangle |$ is the norm of the
pseudo-spin.  
The maximum value for $P_s$
is $a$ and occurs at half-filling; $P_s$ vanishes when the density
reaches $\rho=1\pm b$. At $U=-4$, the numerical MFT values are
$\Delta_d=0.1876$ and $\Delta_p=0.3438$, with $a\approx0.09$ and
$b=1/3$.  Although the maximum value of $P_s$ depends on $U$, we have
found that the value of $b$ is always $1/3$.  Hence, for $\mu=0$, the
mean field $P_s$ always vanishes at fillings $\rho=1\pm 1/3$, {\it
  i.e.} the endpoints of the flat band, in agreement with the QMC
results\cite{footnote4} depicted in Fig.~\ref{Fig:Ps_vs_rho_u-4-8}.
Finally, one has $(n_{\uparrow}-n_{\downarrow})^2= 1-4 \vec{\Delta}
\cdot\vec{\Delta}/3$.  Since, at the mean-field level, $|\langle
\vec{\Delta} \rangle|^2$ is independent of the density within the
range $2/3<\rho<4/3$, this
explains the plateau in Fig.~\ref{Fig:m2_vs_rho_u-4}.
It is remarkable that this behavior is observed in the 
QMC results Fig.~\ref{Fig:m2_vs_rho_u-4} as, in the exact Hamiltonian,
the SU(2) symmetry is only present at $\rho=1$ or $\mu=0$ and not
in the whole range of densities $2/3<\rho<4/3$.

\begin{figure}[t] 
\epsfig{figure=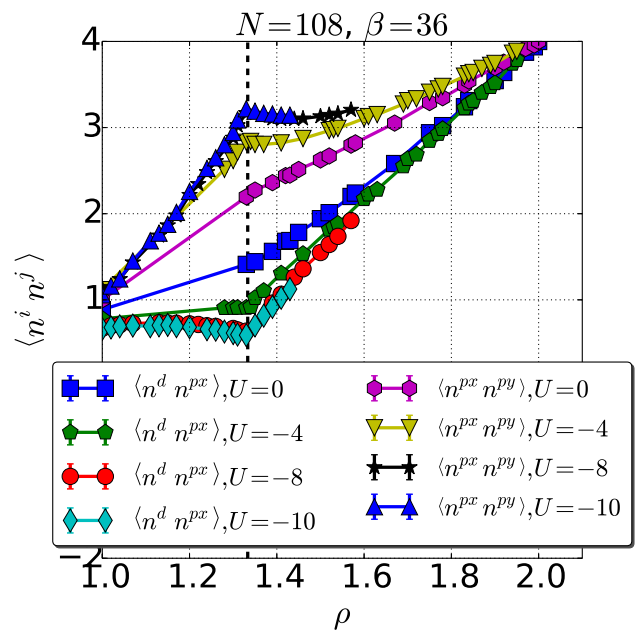,width=8.5cm,angle=0,clip}
\caption{(Color online) Near-neighbor and next near-neighbor
  density-density correlations as functions of $\rho$ for $U=0,-4,-8,
  -10$.  As with the site occupations, these short range density
  correlations exhibit an anomalous decrease even as the total density
  $\rho$ grows.
\label{Fig:ninj_vs_rho_u0-4-8-10}
}
\end{figure}

\subsection{Competition Between Pairing and Charge Order} 
Near-neighbor density-density correlations
(Fig.~\ref{Fig:ninj_vs_rho_u0-4-8-10})  
involving a copper site occupation $\langle n^d n^{px} \rangle$
decrease with increasing $\rho$ in the flat band, reflecting the
transfer of charge to the oxygen sites.  The density correlations
between the two oxygen sites of a unit cell, $\langle n^{px} n^{py}
\rangle$, grow with filling.  The anomalous charge response is
strengthened as the on-site interaction strength $|U|$ becomes larger.

The pair structure factor in the intermediate coupling regime, $U=-4$,
is given in Fig.~\ref{Fig:Ps_vs_rho_u-4-8} (top).  $P_s$ is greatest
when the bands are half-filled, {\it i.e.} for $\rho=1$ and
$\rho=5/3$.  These densities are furthest from the fillings $\rho=2/3$
and $\rho=4/3$ which most favor competing CDW phases
(Fig.~\ref{fig:cdwpatterns2}b,c), and, therefore, vanishing $P_s$.
($P_s$ also vanishes at full filling, $\rho=2$).  Data for $P_s$ do
not show much size dependence for $U=-4t$.  Inset (a) to
Fig.~\ref{Fig:Ps_vs_rho_u-4-8} gives a finite size scaling analysis
\cite{huse88} and supports the existence of pairing LRO in the
thermodynamic limit $1/L \rightarrow 0$, as expected in two dimensions
in the zero temperature limit. For the finite size systems we are
using this limit is reached when the coherence length becomes larger
than the system's size. This result is in agreement with the mean
field results: For $1\le\rho\le4/3$, the mean field pair structure
factor is given by $P_s=P_s^{\text{max}}(1-(3(\rho-1))^2)$, where
$P_s^{\text{max}}=0.09$, in very good agreement with the QMC. For
$4/3\le\rho\le 2$, the mean field results also depict a dome-shape
behavior, with a maximum around the centre, a behavior quite similar
to a single band situation.  The agreement between MFT and DQMC is
less good at larger $|U|$.

\subsection{BEC Within a Flat Band} 
\begin{figure}[t] 
\epsfig{figure=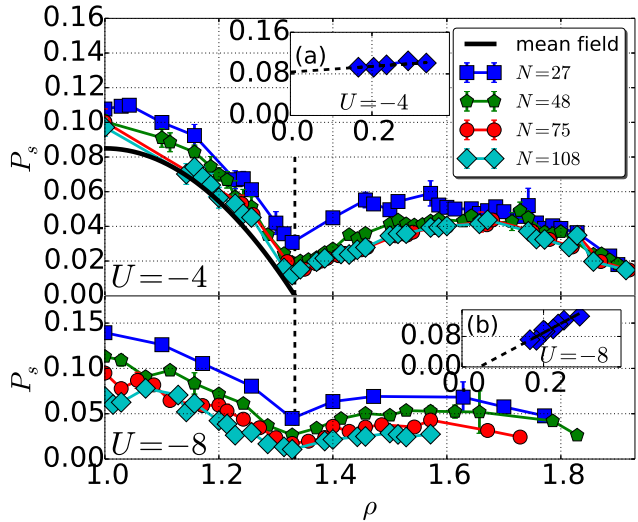,width=8.5cm,angle=0,clip}
\caption{(Color online) Pair structure factor $P_s$ versus $\rho$ for
  $U=-4t$ (top) and $U=-8t$ (bottom). Data for four lattice sizes are
  shown.  $P_s$ has a minimum at $\rho=4/3$ where the superconducting
  phase must compete with 
charge order, and is a maximum when 
$\rho=1$ and $\rho=5/3$.  In the weaker
  coupling case $U=-4t$, at half-filling (inset {\it a}) $P_s$ 
 extrapolates to a non-zero value in the thermodynamic limit.
  However, at 
$U=-8t$, the extrapolation is to zero
  (inset {\it b}).  For $1\le\rho\le4/3$, the mean field pair
  structure factor is given by
  $P_s=P_s^{\text{max}}(1-(3(\rho-1))^2)$, where
  $P_s^{\text{max}}=0.09$, in very good agreement with the QMC
  results. For $4/3\le\rho\le 2$, the mean field calculations depict a
  similar dome-shape behavior, with a maximum around the center, a
  behavior quite similar to a single band situation.
\label{Fig:Ps_vs_rho_u-4-8}
}
\end{figure}

The bottom panel of
Fig.~\ref{Fig:Ps_vs_rho_u-4-8} shows DQMC data at larger $U=-8t$,
approaching the small pair size regime of the AHM.  $P_s$ decreases
much more as the lattice size is increased than for $U=-4t$, and a
finite size scaling analysis (inset (b)) suggests the absence of LRO
at $\beta=36$.  We thus have local pair formation with no clear long
range coherence.  While LRO pair order is likely to develop at a yet
lower energy scale, it is suggestive that it is absent at temperatures
for which superconductivity would be readily visible in dispersing
band geometries like the 2D square lattice: the flat band appears to
be impeding the bosons (locally formed pairs) from forming a BEC.  
For $U=-8t$, pairing LRO is well
established at $\beta=12$ on a square lattice.  In the flat band model
considered here, $\beta=36$ is insufficiently cold.  This factor of
three, or more, reduction in the ordering temperature is much larger
than one would expect simply by the lower coordination number (4 for
the square lattice and 8/3, on average, for the CuO${_2}$ lattice).

\subsection{Comparison with the Kagome lattice}

\begin{figure}[t] 
\epsfig{figure=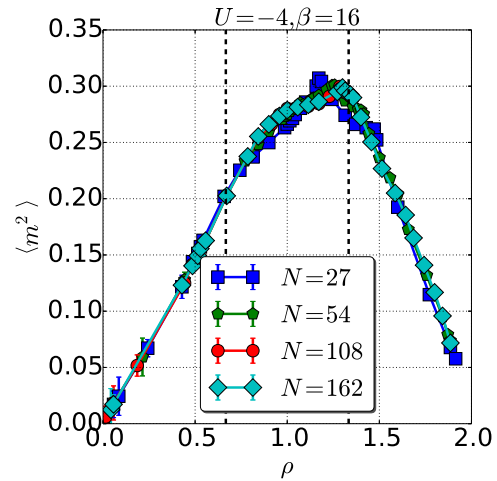,width=6.5cm,angle=0,clip}
\caption{(Color online) Local magnetic moment 
versus $\rho$ for the Kagome
  lattice at $U=-4$. Dashed lines show the non-interacting limit
  between the three bands, the lowest energy band being flat.
   The magnetic moment is approximately proportional to the density $\rho$ in
  the flat band.  Unlike the Lieb case, there is no signature in this 
  quantity as the boundary from the flat band to the second,
  dispersing, band is crossed. 
\label{fig:kagomem2} }
\end{figure}
\begin{figure}[t] 
\epsfig{figure=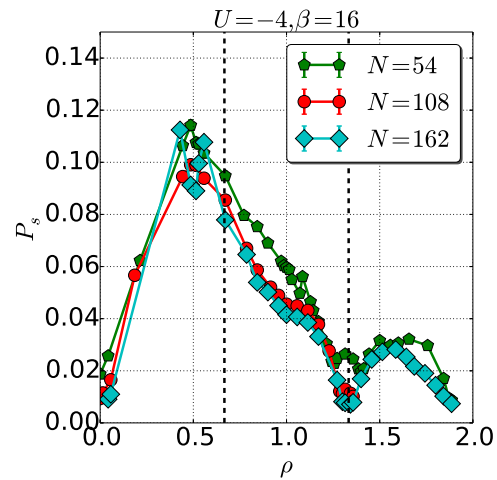,width=6.5cm,angle=0,clip}
\caption{(Color online) Pair structure factor $P_s$ versus $\rho$ for the Kagome
lattice at $U=-4$. As for the magnetic moment (Fig. \ref{fig:kagomem2}), we do 
not observe a peculiar behavior in the flat band region.
\label{fig:kagomeps} }
\end{figure}

In order to assess if these phenomena are generic to all flat band
geometries, we compare our results with the case of the Kagome
lattice.  We choose the sign of the hopping term so that the
dispersionless band is the lowest of the three Kagome bands.  This is
the case of interest to proposed optical lattice experiments on BEC in
Kagome lattices\cite{stamperkurn13}, since condensation occurs
to the lowest energy levels.

Fig.~\ref{fig:kagomem2} and \ref{fig:kagomeps} shows the evolution of the local magnetic moment
$\langle m^2\rangle$  and the pair structure factor $P_s$
for $U=-4$ and different lattice sizes. In the
non-interacting limit, the flat band is occupied for density
$\rho<2/3$.  We see that the novel features that were observed in the
Lieb lattice (figures \ref{Fig:m2_vs_rho_u-4} and
\ref{Fig:Ps_vs_rho_u-4-8}) are no longer present.  Specifically, the
local momemt $\langle m^2\rangle$ is not constant, nor does the pair
structure factor $P_s$ become zero at the edge of the flat band.  This
reflects the absence of competition between CDW and SC order in the
Kagome case.  Moreover, we do not observe a sharp change in the
behavior of either of these observables when the system transitions
from the first to the second band.  There are, however, signatures as
the density takes the Fermi level from the second to the third band,
both of which have non-zero width.  This occurs at $\rho \simeq 4/3$.
The evolution of $P_s$ and $\langle m^2\rangle$ at low density is not
peculiar; indeed it is the one observed in most cases, with $P_s$ and
$\langle m^2\rangle$ roughly proportional to $\rho$.  We observe a
similar behavior in the low or high density limit for the Lieb lattice
(see Figs.~\ref{Fig:m2_vs_rho_u-4} and \ref{Fig:Ps_vs_rho_u-4-8}).

This comparison between the Lieb and Kagome lattices emphasizes that
peculiar behavior, like constant magnetization, observed in the Lieb
case cannot be ascribed solely to a flat band.  In the presence of
interactions, there is no generic evolution of magnetic and pairing
correlations within a flat band.  Instead, other features of the
geometry, such as the presence or absence of frustration, of
particle-hole symmetry, or the existence of distinct types of sites in
the unit cell, also come into play. 

\section{Conclusions} 
Charge and pair correlations in the AHM in the
Lieb lattice, which has a flat band, have been computed.  As the flat
band is filled, the density on the minority (``copper") sites declines
even though the total density grows, 
demonstrating a specific model in which charge transfer signatures are strong.
Such behaviour has
attracted interest in 
the context of cuprate superconductivity and materials like 
BaPbBiO$_3$, where it has been proposed that the
exchange of charge fluctuations can mediate pairing in a way analogous
to the exchange of spin fluctuations \cite{bickers87}.  We have also
presented detailed data on the competition between the pairing and CDW
response as the density is tuned.  At stronger values of the
attraction, 
pairing correlations decrease significantly 
as the lattice size increases, suggesting 
that LRO is inhibited by the flat band
dispersion relative to strongly dispersing bands.  
Finally we compared our results to the Kagome lattice,
another example of a flat band geometry,
and showed that the phenomena observed 
are not generically present in all flat band systems.
Together, these
results characterize `traditional' charge and pairing correlations,
forming a useful context to attempts to explore more exotic
topological phases in flat band systems.

\begin{acknowledgments}
Work supported by the UCOP, by
NNSA-DE-NA0001842-0, by the CNRS(France)-UC Davis EPOCAL LIA joint
research grant and by the CNRS-CQT LIA FSQL.
\end{acknowledgments}

%%%%%%%%%%%%%%%%%%%%%%%%%%%%%%%%%%%%%%%%%%%%%%%%%%%%%%%%%%%%%%%%%%%%%%%%
%% BIBLIOGRAPHY
%%%%%%%%%%%%%%%%%%%%%%%%%%%%%%%%%%%%%%%%%%%%%%%%%%%%%%%%%%%%%%%%%%%%%%%%%%%


\begin{thebibliography}{100}
\bibitem{footnote2} In a three band model of CuO$_2$ planes,
  a site energy difference $\epsilon_p - \epsilon_d > 0$
ensures 
  holes preferentially occupy Cu $d$ orbitals, as occurs in
  the cuprates.  Lieb's theorems do not apply in the presence of such
  a term, which also would create a sign problem in QMC\cite{loh90},
  so we do not include it.
\bibitem{footnote3} Trotter errors from the discretization of the
  inverse temperature are less than a few percent and are smaller than
  statistical error bars in the structure factors.
\bibitem{footnote4} A similar analysis on the square lattice would
  lead to vanishing $b$ coefficient such that $\mu=0$ only corresponds
  to half-filling.
\end{thebibliography}
\end{document}